\documentclass{emulateapj}

\newcommand{\lsun}{L$_{\odot}$}
\shorttitle{Mass ejection from LSPVs}
\shortauthors{Wood \& Nicholls}

\begin{document}


\title{Evidence for mass ejection associated with 
long secondary periods in red giants} 


\author{P. R. Wood and C. P. Nicholls}
\affil{Research School of Astronomy and Astrophysics, Australian National University,\\
Cotter Road, Weston Creek ACT 2611, Australia}
\email {wood@mso.anu.edu.au and nicholls@mso.anu.edu.au}


\begin{abstract}

Approximately 30\% of luminous red giants exhibit a Long Secondary
Period (LSP) of variation in their light curves, in addition to a
shorter primary period of oscillation.  The cause of the LSP has so far
defied explanation: leading possibilities are binarity and a nonradial
mode of oscillation.  Here, large samples of red giants in the Large
Magellanic Cloud both with and without LSPs are examined for evidence of
an 8 or 24 $\mu$m mid-IR excess caused by circumstellar dust.  It is
found that stars with LSPs show a significant mid-IR excess compared to
stars without LSPs.  Furthermore, the near-IR $J$-$K$ color seems
unaffected by the presence of the 24 $\mu$m excess.  These findings
indicate that LSPs cause mass ejection from red giants and that the lost
mass and circumstellar dust is most likely in either a clumpy or a
disk-like configuration.  The underlying cause of the LSP and the mass
ejection remains unknown.

\end{abstract}





\section{Introduction}

When red giants become more luminous than $L \sim 1000$\,\lsun, they
begin to vary with periods of variation which fall on six physically
distinct period-luminosity sequences\footnote{On these sequences, stars on the first
  ascent giant branch (FGB) have slightly longer periods than those on
  the asymptotic giant branch (AGB) at the same luminosity
  \citep{ita04,sosetal07}.  Some authors designate the corresponding FGB
  and AGB sequences as distinct sequences.}
\citep{woo99,ita04,sosetal07,fra08}.  As well as these six sequences for
the most luminous red giants, there is another sequence (known as
sequence-E) extending to lower luminosities and which is known to
consist of close binary systems exhibiting ellipsoidal light
variations \citep{woo99,sos04}.

Among the six luminous sequences all but the longest in period can be
explained by radial pulsation in the fundamental and overtone modes
\citep{woo99}.  The longest period sequence, sequence-D, consists of
stars that exhibit variations with a short period, usually corresponding
to radial pulsation in a low overtone mode, as well as a Long Secondary
Period (LSP).  It is the LSPs that make up sequence-D in the
period-luminosity diagram for variable red giants.  Approximately 30\%
of all luminous red giants have LSPs \citep{woo99,per03,sosetal07,fra08}
so the LSP phenomenon is very common in the late evolution of low mass
stars.  \citet{sosetal07} suggest that LSPs can be seen at very low
amplitude in up to 50\% of luminous red giants.  The LSPs also seem to
occur in red supergiants \citep{sto71}.

Since the period of the LSP is approximately 4 times longer than the
period of the normal-mode radial fundamental mode, the LSP can not be
due to normal-mode radial pulsation.  The most favoured explanations for
the origin of LSPs are binarity and nonradial g-modes
\citep{woo99,hin02,wok04,der06,sos07,hin09}, but these and other explanations
for the LSPs all have significant problems \citep{wok04,nic09}.  For
example, stars exhibiting LSPs have velocity curves which generally are
of similar shape and, in a binary model, this would indicate a favoured
value for the angle of periastron of the orbit rather than the expected
uniform distribution \citep{nic09}.  In addition, the full amplitudes of
the velocity curves are closely concentrated around 3.5 km s$^{-1}$
\citep{nic09} which implies the unlikely situation wherein all these
stars have companions of a similar mass of $\sim$0.09 M$_{\odot}$.  The
problem with the nonradial g-mode explanation for the LSPs is that
g-modes, which have the correct periods, only have substantial
amplitudes in radiative regions and red giants have convective envelopes
with only a thin radiative layer on top \citep{wok04}.  Currently, no
satisfactory explanation for the LSPs exists.

Several of the suggested explanations for the LSPs in red giants involve
circumstellar dust.  The change in visible light associated with the LSP
is typically of a rather irregular nature, and the light variation can
be large, up to a factor of two.  In the binary model, it has been
suggested that the light change could be due to a large dust cloud
orbiting with the companion and obscuring the red giant once per orbit
\citep{woo99}.  Another possible explanation for the LSPs \citep{woo99}
involves semi-periodic dust ejection events such as those predicted by
theoretical models of AGB stars \citep{win94,hof95}.  

Confirmation that dust was involved with the LSP phenomenon would be
obtained if these stars showed a mid-infrared flux excess due to
absorption of stellar light by circumstellar dust followed by
re-radiation in the mid-IR.  Both \citet{hin02} and \citet{ow03}
examined the mid-IR colors of small samples of stars with LSPs in the
solar vicinity using IRAS data.  They found no evidence for a mid-IR
excess that would indicate the presence of unusual amounts of
circumstellar dust.  With the completion of the Spitzer Space Telescope
SAGE survey in the LMC \citep{meix06,blum06}, it is now possible to
search for the presence of a mid-IR excess in large samples of stars
with LSPs, good contemporary light curves and a known distance.  We now
describe such a search.

\section{Selection of samples of variables}\label{samp_sec}

\begin{figure}
\epsscale{1.00}
\plotone{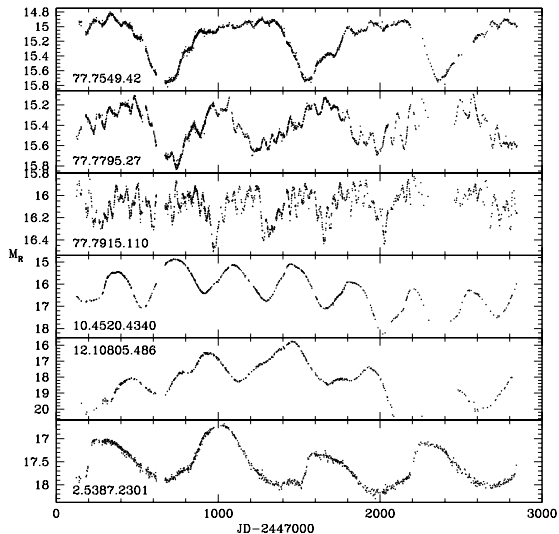}
\caption{Typical light curves of sequence-D stars (top 3 panels) and of
  high mass loss rate stars removed from the \citet{fra08} list of
  sequence-D stars (lower 3 panels).}
\label{lcs}
\end{figure}

The sample of variable stars examined here is taken from \citet{fra08} who
classified the luminous red giant variables in the LMC according to
whether they belonged to sequence-D or sequences 1--4\footnote{Other authors use
sequences C, C$'$, B and A in place of sequences 1--4, respectively, of
\citet{fra08}.  In addition, \citet{sosetal07} find a shorter period sequence
that would be sequence 5 in the notation of \citet{fra08}.}.  The sequence-D
stars exhibit light variations at both a primary period, which is
usually on sequence-2 or sequence-3, and a Long Secondary Period, which
falls on sequence-D.  Examples of the light curves of typical sequence-D
stars are shown in the top three panels of Figure~\ref{lcs}.

The sequence-D sample of \citet{fra08} was refined in a number of
ways.  To be retained in our sample of sequence-D stars, a star was
required to show both the quasi-periodic LSP as well as a shorter
primary period.  In an initial examination of the light curves of the
sequence-D stars in the catalog of \citet{fra08}, we found stars which
have light curves that show a smooth, large amplitude variation as
well as a variation of mean magnitude over very long timescales.
Examples of their light curves are shown in the lower three panels of
Figure~\ref{lcs}.  These are stars with large-amplitude, Mira-like
pulsation, high mass loss rates and thick circumstellar dust shells
and they belong on sequence-1.  The dust shells cause them to appear
fainter in the visible and near-IR part of the spectrum than they
would appear without the dust shell.  In the ($K$,$\log P$) plane,
they thus lie below sequence-1, and they can take a position near or
on sequence-D \citep[e.g.][]{woo03}.  In order to decrease the number
of light curves to be examined to test for genuine sequence-D
characteristics, we restricted our sample of sequence-D stars to have
a value for the LSP falling on the main body of the sequence, defined
by $2.1 < \log P - (14-K)/4.2 < 2.4$.  This eliminates considerable
numbers of the dusty, Mira-like stars assigned to sequence-D by
\citet{fra08}, as well as some sequence-E stars.  The light curves of
all remaining stars within the main body of sequence-D and with $K <
12$ were inspected and the dusty, Mira-like stars and sequence-E stars
were removed.  We limited our detailed examination of light curves
to $K < 12$ as it is these brighter stars that are used later in the
paper for the comparison of mid-IR properties.  A small number of
other stars that did not have the required sequence-D light curve
shape were also removed: these were normal semi-regular variables
without obvious LSPs that are typical of sequences 2 and 3, and
several R Coronae Borealis stars.  In total, we removed 105 stars from
the original sequence-D sample.  Finally, about 5\% of stars appeared
twice in the lists of \citet{fra08} under different MACHO names
(because they were in overlapping parts of different MACHO fields):
only one entry was retained for these stars.

Our aim is to see if unusual amounts of circumstellar dust are
associated with stars exhibiting LSPs.  In order to make such a test, we
need a sample of stars that are identical to the stars with LSPs except
that they do not have LSPs.  This comparable sample is the combined
group of stars on sequence-2 and sequence-3 i.e.~stars with an
oscillation similar to the primary oscillation period of the LSPVs.
Hereinafter, we will refer to the variable stars on sequence-D as LSPVs and the stars
on sequence-2 and sequence-3 as SRVs (semi-regular variables).

Looking at the light curves of the LSPVs and SRVs, it is clear that some
variations are of low amplitude and therefore the periods, and hence the
variability sequence, are not well determined.  \citet{fra08} determined
periods by two methods - a Fourier analysis and a SuperSmoother method.
In order to get samples of LSPVs and SRVs with reliable sequence
determinations, we chose a subsample of the LSPVs and SRVs in which the
SuperSmoother and Fourier period agreed to 5\%.  We call these the P5
samples.  The stars in the P5 samples tend to have larger amplitudes and
more regular variations than the remaining stars in the complete
samples.  The amplitude distribution of the two samples is shown later,
in Section~\ref{ampd}.  For the stars with $K < 12$ that are examined in
detail in this paper, there are 4807 LSPVs and 4568 SRVs in the complete 
samples and 1568 LSPVs and 971 SRVs in the P5 samples.

\section{Testing for circumstellar dust using mid-IR fluxes from the SAGE survey}
   
Since cool circumstellar dust emits preferentially in the mid-IR
(longward of $\sim$5\,$\mu$m) while the stellar photosphere emits mainly
in the near-IR (shortward of $\sim$3\,$\mu$m), a large ratio of mid-IR
flux to near-IR flux can be used as an indicator of circumstellar dust.
The $K$-$[24]$ color is a sensitive indicator of cool circumstellar dust,
especially very cool dust with a temperature of $\sim$200\,K, while the
$K$-$[8]$ color would be more sensitive to slightly warmer dust with $T
\sim 400$\,K.  A problem with the $K$-$[24]$ color is that the SAGE survey
only detected a 24\,$\mu$m flux from a small fraction of the bright red
giants in the LMC.  On the other hand, and 8\,$\mu$m flux was detected
for most bright red giants.  We now discuss both of these options.

\subsection{The $K$-$[8]$ color}

\begin{figure}
\epsscale{1.0}
\plotone{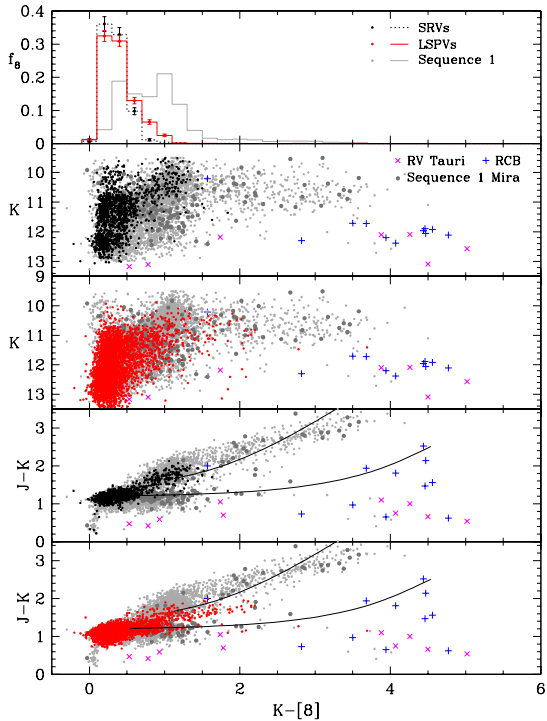}
\caption{$K$ and $J$-$K$ plotted against $K$-$[8]$ for the P5 samples of
  LSPVs (small red points, panels 1 and 3 from the bottom) and SRVs (small black points,
  panels 2 and 4 from the bottom).  Stars on sequence-1 are shown as small grey 
  points.  The large amplitude Mira variables belonging
  to sequence-1 are shown as larger points that are a darker shade of
  grey.  The magenta crosses are the 10 RV Tauri stars in the LMC
  studied by \citet{alcock98} that have $J$, $K$ and $[8]$ magnitudes
  while the blue plus signs are the 11 RCB stars in the LMC studied by
  \citet{alcock01}.  The lines in the bottom two panels are model AGB
  stars with spherical dust shells of increasing optical depth (see
  text).  The top panel shows the fraction of the O-rich ($J$-$K$ $<$ 1.4) 
  SRVs and LSPVs stars brighter than $K
  = 12$ which lie in $K$-$[8]$ color bins of width 0.2 magnitudes:
  one-sigma error bars are given.  Also shown in the top panel is the 
  distribution of sequence-1 stars, both C and O-rich.}
\label{mir_colour8}
\end{figure}

In Figure~\ref{mir_colour8} we show $K$ and $J$-$K$ plotted against
$K$-$[8]$ for the LSPVs and the SRVs in the P5 samples.  Also shown are
stars on sequence-1: these are fundamental mode radial pulsators and
they include the Mira variables and large amplitude, high mass loss
rate, dust enshrouded long period variables.

It is well known that, in the LMC, red giant stars with $J$-$K \ga 1.4$
are mostly carbon stars and those with $J$-$K \la 1.4$ are oxygen-rich M
or K stars \citep{ch03}.  The separation of the C- and O-rich stars is
seen in the ($K$-$[8]$,$J$-$K$) diagrams in the lower two panels of
Figure~\ref{mir_colour8}: the separation is seen most easily for the
grey points of sequence-1.  These diagrams also show that there is a long
tail of sequence-1 stars extending from the clump of C stars around
($K$-$[8]$,$J$-$K$) = (1.1,1.8) to redder $K$-$[8]$ and redder $J$-$K$
colors.  This tail contains many C-rich Mira variables, defined as
sequence-1 variables with a MACHO blue amplitude greater than 2.5
magnitudes.  However, the Mira variables are not confined to this region
of the ($K$-$[8]$,$J$-$K$) diagram.  There are even more of them lower in
the diagram with $J$-$K$ $<$ 1.4 and $K$-$[8]$ $<$ 1.5: these are O-rich
Mira variables.

Mira variables are AGB stars of relatively high mass loss rate with
roughly spherical shells, perhaps with some clumpiness (see
\citealt{olo04} for a review of the evidence for clumpiness in Mira
winds).  In order to get some idea of the positions of stars with
spherical mass loss shells in the ($K$-$[8]$,$J$-$K$) diagram, models of
such stars have been made with the code DUSTY \citep{ive99} which is
designed to model the spherical dust envelopes of mass-losing AGB stars.
Two series of models were made, one for C stars and one for O-rich
stars.  The C star models had a central star which was assumed to be a
blackbody of temperature 2200 K and the grains in the stellar wind
emanating from the central star were assumed to be 50\% amorphous carbon
and 50\% SiC.  The O-rich models had a central blackbody with $T$ = 2700
K and warm silicate dust in the stellar wind.  In both sets of models,
the inner temperature of the dust shell was assumed to be 1000 K, and
the model series had a range of V-band optical depths from 0.1 to 10,
corresponding to mass loss rates of approximately 2$\times$10$^{-7}$ to
7$\times$10$^{-6}$ M$_{\odot}$ yr$^{-1}$ for a star with $L =
5000$\,L$_{\odot}$.  These are simple approximations that do not allow
for variation of the stellar temperature, inner boundary temperature or
grain composition as the stars evolve.  They are meant to provide a
general idea of how mass-losing AGB stars evolve in the
($K$-$[8]$,$J$-$K$) diagram rather than a precise prediction.

The upper curve in the lower two panels of Figure~\ref{mir_colour8}
represents the sequence of model C stars with increasing mass loss rate.
This corresponds, within the simple model approximations used here, to
the tail of sequence-1 stars in the upper parts of the ($K$-$[8]$,$J$-$K$)
diagrams.  This tail thus represents the evolutionary sequence of C
stars with spherical mass loss shells of increasing mass loss rate (see
\citealt{gro07} for detailed mass loss rate estimates for these stars).
The lower curve in the bottom two panels of Figure~\ref{mir_colour8} is
the equivalent curve for O-rich stars.  It can be seen that many O-rich
Miras fall on this curve.

The position of the high mass loss rate AGB stars with spherical shells
contrasts with the position of the RV Tauri stars shown in
Figure~\ref{mir_colour8}.  The latter can have very large $K$-$[8]$ colors
but their $J$-$K$ colors are similar to the colors of stars with unobscured
photospheres.  Since RV Tauri stars are luminous post-AGB
stars containing a dusty circumbinary disk which radiates in the mid-IR
\citep{deRuyter05}, a low $J$-$K$ and a large $K$-$[8]$ is likely to
indicate a star with a dusty disk.  This picture is consistent with the
$J$-$K$ colors of the R Coronae Borealis (RCB) stars, which are all
C-rich.  The dust associated with them is in neither a disk nor a
spherical shell but arises from the ejection of puffs of dust in random
directions, leading to a highly non-spherical dust distribution
\citep{cla96}. The $J$-$K$ colors of the RCB stars scatter between the
$J$-$K$ colors of the RV Tauri stars and the high mass loss rate C-rich
spherical shell sources, as expected in this model since the line of
sight to some RCB stars would contain dust clouds while in other stars
the photosphere would be seen unobscured.

The considerations above allow us to interpret the positions of the SRVs
and LSPVs in Figure~\ref{mir_colour8}.  The great majority of the
oxygen-rich SRVs ($J$-$K < 1.4$) have $K$-$[8]$ $<$ 0.7, indicating very
little mid-IR excess and low mass loss rates.  Similarly, the C-rich
SRVs lie mostly in the domain of the C stars with low mass loss rates on
sequence-1.  This contrasts with the LSPVs, both C- and O-rich, many of
which appear to show some evidence for a $K$-$[8]$ excess.  
For the C stars, the large fraction of LSPVs with a $K$-$[8]$ excess 
can be clearly seen by comparing the lower
two panels of the figure.  The top panel
of Figure~\ref{mir_colour8} shows the $K$-$[8]$ color distribution of the
O-rich SRVs and LSPVs more quantitatively: the larger fraction of LSPVs
with $K$-$[8]$ $>$ 0.7 is clearly seen (11\% of LSPVs have $K$-$[8]$
$>$ 0.7 compared to 2.3\% of SRVs).  A two sample K-S test gives a
probability less than $3.8\times10^{-7}$ that the SRVs and the LSPVs
in the histograms come from the same underlying distribution i.e. the
higher fraction of LSPVs with $K$-$[8]$ $>$ 0.7 is extremely unlikely to
be a statistical artifact.  Note that in the histograms shown in
Figure~\ref{mir_colour8}, and in subsequent figures, we have restricted the
samples of stars to those with $K < 12$ since most of the stars that
have detectable 8 and 24\,$\mu$m fluxes have $K < 12$.  In addition,
first ascent giant branch stars are excluded by the restriction $K <
12$.  At 8\,$\mu$m, we find that 99.4\% of LSPVs (1561 stars), 99.8\% of
SRVs (969 stars) and 99.5\% of sequence-1 variables (3964 stars) with $K
< 12$ were detected.

There is one notable feature of the ($K$-$[8]$,$J$-$K$) diagram for LSPVs.
The reddest C-rich LSPVs have $J$-$K$ colors that are similar to those of
the unreddened C stars, and they are distinctly less than the $J$-$K$
colors of the C-rich circumstellar shell sources in sequence-1 which
have the same $K$-$[8]$ color.  There are also a few O-rich LSPVs with
large $K$-$[8]$ colors that fall near the RV Tauri stars.  This suggests
that mass loss from LSPVs is not spherically symmetric and that the dust
may be in a circumstellar disk or some other non-spherical distribution.
These results become much clearer when using the 24 $\mu$m measurements
from the SAGE survey, as we will now show.

\subsection{The $K$-$[24]$ color}

\begin{figure}
\epsscale{1.0}
\plotone{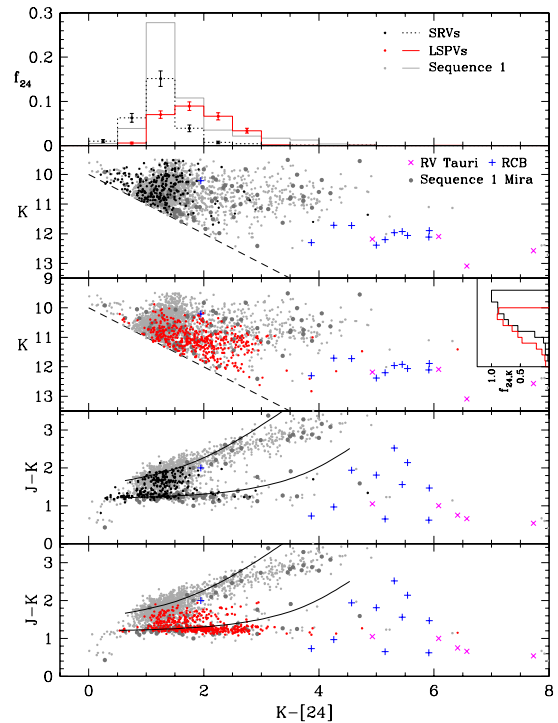}
\caption{The same as Figure~\ref{mir_colour8} but using the $K$-$[24]$
  color rather than the $K$-$[8]$ color for the P5 samples.  The dashed
  line marks the 24\,$\mu$m magnitude limit ($[24]$ = 10) of the SAGE
  survey.  The inset in the third panel shows the fraction of the P5 sample
  of LSPVs (red histogram) and SRVs (black histogram) detected at
  24\,$\mu$m as a function of K magnitude.  The histograms of LSPVs and
  SRVs in the top panel include both C and O-rich stars.}
\label{mir_colour24good}
\end{figure}

Figure~\ref{mir_colour24good} shows a plot similar to
Figure~\ref{mir_colour8} but using the $K$-$[24]$ color.  It is
immediately obvious from comparing Figures~\ref{mir_colour8} and
\ref{mir_colour24good} that only a small fraction of the sample was
detected at 24\,$\mu$m.  In the P5 samples, for objects brighter than $K
= 12$, 27\% of LSPVs (424 stars) and 28\% of SRVs (269 stars) were
detected at 24\,$\mu$m whereas essentially all stars were detected at
8\,$\mu$m.  For the sequence-1 variables, 52\% (2077 stars) brighter
than $K = 12$ were detected at 24\,$\mu$m.

Comparing the LSPV and SRV samples in the lower four panels of
Figure~\ref{mir_colour24good}, it appears that the LSPVs are much more
likely to show a $K$-$[24]$ excess than the SRVs.  This is confirmed by
the histogram of $K$-$[24]$ colors in the top panel of the figure where it
can be seen that a significantly larger fraction of LSPVs with $K < 12$
have $K$-$[24]$ $>$ 1.5 (72\% of LSPVs have $K$-$[24]$ $>$ 1.5 compared to
19\% of SRVs).  A two sample K-S test gives a probability less 
than $2.0\times10^{-42}$ that the SRVs and the LSPVs in the histograms come
from the same underlying distribution.

Given the stronger mid-IR excess exhibited by the LSPVs, one would
expect a larger fraction of LSPVs to be detected at 24\,$\mu$m than SRVs
yet, as noted above, similar fractions of LSPVs (27\%) and SRVs (28\%)
were detected.  This is explained by the inset histogram in the middle
panel of Figure~\ref{mir_colour24good} which shows the detection
fraction as a function of $K$ magnitude.  It is seen that the SRVs
extend to brighter $K$ magnitudes than the LSPVs and that there is
almost 100\% detection of these brighter objects even though they do
not have large $K$-$[24]$ values.  Most importantly, at all $K$ magnitudes
where the LSPV and SRV samples both exist, the LSPV detection fraction
is higher than the SRV detection fraction.

The C stars and the O-rich stars in the lower two panels of
Figure~\ref{mir_colour24good} fall in distinct groups separated by the
line $J$-$K$ = 1.4, the C stars lying above this line.  As with
Figure~\ref{mir_colour8}, the $J$-$K$ colors of the C-rich LSPVs in
Figure~\ref{mir_colour24good} are smaller than those of the C-stars on
sequence-1 which have high mass loss rates in spherical winds.  The
picture for the O-rich stars is less clear.  On the one hand, there is a
very large group of O-rich LSPVs with $J$-$K$ $<$ 1.4 but red $K$-$[24]$
colors which fall below the line of DUSTY models of increasing mass loss
rate in spherical shells.  This group of O-rich LSPVs seems to follow the
sequence occupied by the RV Tauri stars with circumstellar disks.  On
the other hand, some sequence-1 Mira variables also follow this sequence
while others have larger $J$-$K$ values as suggested by the spherically
symmetric models.  It is not clear whether this diversity of behaviour is because the
winds in Miras are not all spherically symmetric or if it is because the
models are too simplistic\footnote{The O-rich models of \citet{gro06} follow the 
same general behaviour as the DUSTY models shown here but they tend to be 0.5
to 1 mag redder in $K$-$[24]$ for the higher mass loss rates.}.  
However, the reddest LSPVs with $K$-$[24]$
$>$ 4 appear similar to the RV Tauri stars and they almost certainly
have a non-spherically symmetric dust distribution.  Overall, these
results indicate that not only do the LSPVs produce more dust than
equivalent SRVs without LSPs, but this dust appears to be in a
non-spherical distribution about the star, possibly in a circumbinary
disk.

\begin{figure}
\epsscale{1.0}
\plotone{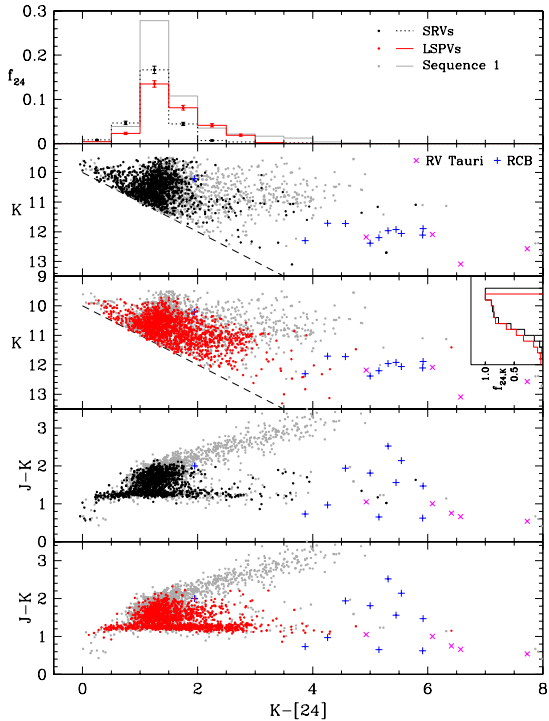}
\caption{The same as Figure~\ref{mir_colour24good} but for stars
detected at 24\,$\mu$m in the larger, complete samples.
}
\label{mir_colour24all}
\end{figure}

The numbers of stars in the P5 samples are moderate.  However, the
results shown above are confirmed by the larger number of stars in the
complete samples, as shown in Figure~\ref{mir_colour24all}.  In the
complete samples, for objects brighter than $K = 12$, 31\% of LSPVs
(1482 stars) and 28\% of SRVs (1285 stars) were detected at 24 $\mu$m.
These fractions are similar to the detection fractions in the P5 samples.

\begin{figure}
\epsscale{1.0}
\plotone{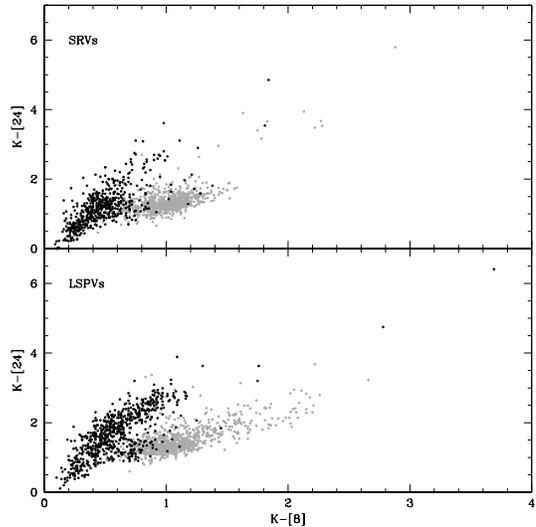}
\caption{
$K$-$[24]$ plotted against $K$-$[8]$ for the stars that have $K < 12$
in the complete samples of LSPVs and SRVs.  The black points are
stars with $J$-$K < 1.4$ (O-rich star candidates) while the grey
points are stars with $J$-$K \ge 1.4$ (C star candidates).
}
\label{two_colour_all}
\end{figure}

One question that arises is whether the stars that show a $K$-$[24]$
excess are also the stars that show a $K$-$[8]$ excess.  We note that
there are only 3 stars in the combined complete samples with a
24\,$\mu$m detection but not an 8\,$\mu$m detection, so that it is
possible to compare the $K$-$[24]$ excess with the $K$-$[8]$ excess
for essentially all stars.  This is done in
Figure~\ref{two_colour_all} where $K$-$[24]$ is plotted against
$K$-$[8]$ for all stars with $K < 12$ in the complete samples.  There
are clearly two sequences in Figure~\ref{two_colour_all}, one for C
stars and one for O-rich stars.  Both sequences show a correlated
increase in $K$-$[8]$ color with $K$-$[24]$ color.  Thus stars with a
$K$-$[24]$ excess do also show a $K$-$[8]$ excess.  Note that the
LSPVs have a higher fraction of red stars than the SRVs (as already
demonstrated in Figures~\ref{mir_colour8}--\ref{mir_colour24all}).

\subsection{The effect of amplitude on $K$-$[24]$ color}\label{ampd}

It has been shown above that the existence of a Long Secondary Period in
a red giant seems to induce extra circumstellar dust formation.  In this
situation, it might be expected that the infrared excess caused by dust
would depend on the light amplitude of the LSP.  We now examine this
suggestion.

\begin{figure}
\epsscale{1.0}
\plotone{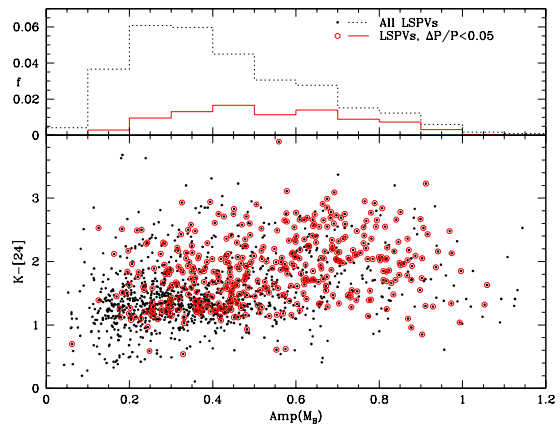}
\caption{The $K$-$[24]$ color plotted against the MACHO blue amplitude of
  the LSP for LSPVs with $K < 12$ (bottom panel).  Black dots show stars
  in the complete sample while stars also in the P5 sample are
  surrounded by red circles.  The histogram in the top panel shows the fraction of all
  LSPVs which have $K < 12$ and a detected 24\,$\mu$m flux 
  in the P5 sample (red line) and the complete sample
  (black dotted line).
}
\label{amphist}
\end{figure}

Figure~\ref{amphist} shows the $K$-$[24]$ color plotted against the MACHO
blue amplitude for LSPVs in the P5 and complete samples.  We used the
MACHO blue amplitude rather than the MACHO red amplitude as the phase
coverage is generally better in blue than in red.  

There is a slight increase of the $K$-$[24]$ color with MACHO blue
amplitude.  A least squares fit gives $K-{\rm [24]} = 0.73 Amp(M_{B})
+1.47$ for the complete sample and $K-{\rm [24]} = 0.90 Amp(M_{B})
+1.20$ for the P5 sample.  In both cases, the {\it rms} scatter of the
$K$-$[24]$ color about the fit is 0.50 magnitudes.  Given this large
scatter, a wide range in $K$-$[24]$ color can be found at any given light
amplitude.  We conclude that the light amplitude does not seem to be an
important factor in determining the amount of mass loss and the mid-IR
excess of LSPVs.

The top panel of Figure~\ref{amphist} indicates that the LSPVs
that are in the P5 sample have larger average amplitudes than those in
the complete sample.  This confirms our assertion in
Section~\ref{samp_sec} .

It would be interesting to see if there was any correlation between the
velocity amplitude of the LSP given in \citet{nic09} and the $K$-$[24]$
color.  Note that the radial velocity amplitude associated with the LSPs
does not show any correlation with light amplitude \citep{nic09}.  There
are 16 LSPVs in the group of stars studied by \citet{nic09} that have
$K$-$[8]$ colors and 6 with $K$-$[24]$ colors.  A plot of the $K$-$[8]$ and
$K$-$[24]$ colors against velocity amplitude for these small samples of
stars did not show any particular correlation.

\section{Summary and conclusions}

We have shown that luminous red giant stars that exhibit Long Secondary
Periods have larger mid-IR fluxes than similar stars without LSPs.  This
suggests that the LSP induces additional mass loss from the red giant,
with consequent dust formation and an increase in the mid-IR flux.  The
mid-IR flux excess is only weakly dependent on the amplitude of the light
variation of the LSP.  A comparison of the near-IR $J$-$K$ color with
the mid-IR $K$-$[24]$ and $K$-$[8]$ colors indicates that the dust is not in
a spherically symmetric distribution, and is perhaps in a disk.

Although dust is associated with the LSP phenomenon, it is still not
clear what causes the LSP in the first place.  In a binary model, where
the velocity curves indicate predominantly eccentric orbits
\citep{nic09}, mass transfer from the red giant near periastron
could lead to a circumbinary disk.  A non-radial pulsation mode of low
degree could also produce non-spherical mass loss and a non-spherical
dust distribution.  The semi-periodic dust ejection events predicted by
\citet{win94} and \citet{hof95} produce, and indeed require,
circumstellar dust.  However, this phenomenon has not been found to
occur in theoretical models at the low luminosities where many of the
LSPs are observed (even below the tip of the FGB).  Furthermore, it is
also not clear why the photospheric velocity should vary with the LSP
for this purely circumstellar process.

One additional characteristic of LSPVs is that they have a chromosphere
which varies with the LSP \citep{wok04}.  It is likely that both the
chromosphere and the excess circumstellar dust are manifestations of the
influence of the LSP phenomenon on matter above the stellar photosphere.
Magnetic effects are a possible source of the chromosphere, but a search
for magnetic fields in two solar-vicinity LSPVs has put an upper limit
of 100 Gauss on magnetic fields covering more than about 10\% of the
surface of these stars \citep{wmwn09}.  Currently there is no evidence
that magnetic fields are the source of the chromosphere.

In summary, we have shown here that the LSP phenomenon produces excess
circumstellar dust compared to stars without LSPs.  Unfortunately,
since all the postulated models for the LSP are potentially capable of
causing mass ejection, the present detection of excess dust around LSPVs
does not help us distinguish between the various models.


%% Included in this acknowledgments section are examples of the
%% AASTeX hypertext markup commands. Use \url without the optional [HREF]
%% argument when you want to print the url directly in the text. Otherwise,
%% use either \url or \anchor, with the HREF as the first argument and the
%% text to be printed in the second.

\acknowledgments 
We are grateful to the SAGE team for
creation of the SAGE legacy data and to the authors of DUSTY
for making their code publicly available.  
This paper utilizes public domain data originally
obtained by the MACHO Project, whose work was performed under the joint
auspices of the U.S. Department of Energy, National Nuclear Security
Administration by the University of California, Lawrence Livermore
National Laboratory under contract No. W-7405-Eng-48, the National
Science Foundation through the Center for Particle Astrophysics of the
University of California under cooperative agreement AST-8809616, and
the Mount Stromlo and Siding Spring Observatory, part of the Australian
National University.

\end{document}